\documentclass[13pt]{iopart}

\usepackage{iopams}
\expandafter\let\csname equation*\endcsname\relax
\expandafter\let\csname endequation*\endcsname\relax
\usepackage{amsmath}
\usepackage{graphicx}\usepackage{multicol}
\usepackage{xcolor}
\usepackage{xcolor}
\usepackage{braket}\usepackage{bm}
\usepackage{caption}
\usepackage{subcaption}
\usepackage{amssymb}\usepackage{amsmath}
\usepackage{bm}\usepackage{float}\usepackage{xcolor}
\usepackage{booktabs}\usepackage{multicol}
\usepackage{graphicx}
\usepackage[margin=0.9in]{geometry}\usepackage{subcaption}
\begin{document}
\title[Kabir et al.,]{Re-analysis of Temperature Dependent Neutron Capture Rates and Stellar $\beta^{-}$ decay Rates of $^{95-98}$Mo.}
\author{Abdul Kabir$^{1}$, Jameel-Un Nabi$^{2}$, Muhammad Tahir$^{2}$, Abdul Muneem$^{3}$ and Zain Ul Abideen$^{1}$.}
\address{$^{1}${Space and Astrophysics Research lab, National Centre of GIS and Space Applications, Department of Space Science, Institute of Space Technology, Islamabad 44000, Pakistan.}}
\address{$^{2}${University of Wah, Quaid Avenue, Wah Cantt 47040, Punjab, Pakistan.}}
\address{$^{3}${Faculty of Engineering Sciences, GIK Institute of Engineering
		Sciences and Technology, Topi 23640, Khyber Pakhtunkhwa,
		Pakistan.}}
\ead{kabirkhanak1@gmail.com}
\vspace{10pt}
\begin{abstract}
The neutron capture rates and temperature dependent stellar beta decay rates of Mo isotopes are investigated within the framework of the statistical code (Talys v1.96) and proton-neutron quasi particle random phase approximation (pn-QRPA) models. The Maxwellian average cross-section  (MACS) and neutron capture rates for $^{95-98}$Mo(n,$\gamma$)$^{96-99}$Mo radiative capture processes are analyzed within the framework of statistical code Talys v1.96 based on the  phenomenological nuclear level density (NLD) parameters and gamma strength functions (GSFs). The current model-based computed data for MACS provide a good comparison with the existing measured data. The temperature-dependent stellar $\beta$ -decay rates of $^{95-98}$Mo are investigated at different densities within the framework of the pn-QRPA model. For the considered temperature range, the neutron capture rates were found to be higher, by orders of magnitude, than the stellar $\beta$ rates.
\end{abstract}

\vspace{2pc}
\ioptwocol
\maketitle

\section{Introduction}
The energy production in stars \cite{Bethe}, the associated nucleosynthesis \cite{Burbidge,Cameron} and supernova
explosion dynamics \cite{Baade} are still not fully understood. To date, these processes have been extensively
investigated by astrophysicists in an attempt to understand
how our universe works. Weak-interaction mediated rates play a crucial role during the nucleosynthesis process and the evolution of the core of massive stars.

Nuclei, more massive than iron, are mostly formed by neutron capture reactions, which are called after their widely distinct time scales as the $s$-process (slow neutron capture process) and the $r$-process (rapid neutron capture process) \cite{Burbidge,Cameron}. The $s$-process in the present model of stellar nucleosynthesis is more challenging than previously believed~\cite{Kappeler11}. The elements from Fe to Zr (A$\leq$90) are formed by the so-called weak component, which is driven by neutrons produced through the $^{22}$Ne($\alpha$,n)$^{25}$Mg reaction in massive stars. Thus, a stellar model to describe the weak $s$-process requires accurate and reliable Maxwellian average cross-section (MACS) at $s$-process relevant temperatures. The main $s$-process operates in the AGB stars and produces the heavier isotopes up to the lead-bismuth region. Because of the breakthroughs in astronomical observations and stellar modeling, accurate assessments of the $s$-process have become incredibly influential~\cite{Lugaro}. The current stellar model suggests that the neutron production for the weak $s$-process occurs in two different burning stages: the He core burning and the Carbon shell burning~\cite{Raiteri}. The burning temperatures of the two stages are not the same (T$_9 \sim $ 0.3 for the He core burning and T$_9 \sim$ 1 for the carbon shell burning, where T$_9$ is the temperature in units of 10$^{9}$ K). Hence, the $s$-process takes place when $0.3<\rm T_9<1$ and at neutron density of $N_n$$\approx$ $10^8$~cm$^{-3}$~\cite{Kappeler90}.  The $s$-process has evolved from a simple explanation of the abundance distribution of elements in the solar system to a more comprehensive  description that takes into account the general features of stellar and galactic dynamics. As a result of such advancements, the $s$-process has become a powerful tool for examining the evolution of red giant stars~\cite{Busso}.

Here, we focus the reader's attention towards the race between $^{95-98}$Mo(n,$\gamma$)$^{96-99}$Mo and $^{95-99}$Mo$ \rightarrow$ $^{95-99}$Tc + $e$ + $\nu_e$  processes within the framework of Talys v1.96 and the pn-QRPA approach. Substantial progress has been achieved in the analysis of Mo nuclei from the  astrophysical perspectives. Advances in nuclear modeling and observational techniques have allowed researchers to gain a better understanding of the nucleosynthesis of heavy elements in stars. The MACS of $^{95-98}$Mo are crucial quantities in determining the rates. 
Neutron interaction cross-sections with Molybdenum play a crucial role in diverse scientific and technological applications, from nuclear astrophysics to the operation of nuclear power plants. Mo isotopes are identified as a contaminant in pre-solar silicon carbide grains and play a critical role in the nucleosynthesis processes occurring in the AGB stars \cite{liu}. It is suggested that low uncertainty in neutron capture cross-section of Mo isotopes is highly desirable for accurate mapping of nucleosynthesis in low-mass AGB stars \cite{cescutti}. In the past, several authors have  studied neutron capture cross-sections of Mo isotopes at different energy ranges,  theoretically and experimentally \cite{Saumi,pomerance,kapchigashev,stupegia}. Saumi et al. \cite{Saumi} studied the MACS for the nuclei participating in the $s$-process and the $p$-process nucleosynthesis in and around the N= 50 closed neutron shell. They have constructed a microscopic optical-model potential by the folding DDM3Y nucleon-nucleon interaction with the radial matter density of the target, obtained from relativistic mean-field (RMF). Their computed MACS at kT=30 keV for the $^{95-98}$Mo(n,$\gamma$)$^{96-99}$Mo were 212~mb, 113~mb, 299~mb and 73.8~mb, respectively. Neutron capture cross-sections of seven stable isotopes of Mo have been measured using a 40~m station of the Oak Ridge Electron Linear Accelerator (ORELA) in the energy range (3$<E_{n}<$90)~keV \cite{musgrove}. Recently, Ref.~\cite{duhamel} measured the capture cross-section and transmission of natural Mo using an accurate neutron-nucleus reaction measurement instrument (ANNRI) situated in the material life and science facility (MLF) at J-PARC. Measurements of capture and total cross-sections for isotopes $^{94,95,96}$Mo have been conducted at neutron time-of-flight facilities, n\_TOF at CERN and GELINA at JRC-Geel, using samples with an enrichment exceeding 95\%  for each isotope  \cite{mucciola}. The transmission parameters obtained in the experiment have been used to validate the resonance parameter files for Mo isotopes. Massimi \textit{et al}. \cite{massimi} measured the neutron capture cross-section in EAR1 and EAR2 at the n\_TOF facility to reduce the uncertainty in the presently known data for the stable isotopes $^{94,95,96}$ Mo. $^{98}$Mo is the stable and most abundant isotope with an abundance of 24.13\%. Understanding the thermal neutron cross-section and the resonance integral for the $^{98}$Mo(n,$\gamma$)$^{99}$Mo reaction is crucial due to the utilization of neutron activation cross-section data in the production of $^{99}$Mo and its potential applications in various studies related to neutron-matter interactions \cite{Ref}.Existing literature contains numerous experimental and evaluated datasets on the thermal neutron capture cross-sections and resonance integrals for the $^{98}$Mo(n,$\gamma$)$^{99}$Mo reaction. The authors in Ref.~\cite{uddin}  studied the cross-section of $^{98}$Mo(n,$\gamma$)$^{99}$Mo using monochromatic thermal neutron beam at low energies. The obtained cross-sections are  116$\pm$7 mb and 91$\pm$5 mb at energies 0.0334 eV and 0.0536 eV, respectively, agrees with the ENDF/B-VII data files and JENDL-4. The formation cross-section of $^{99}$Mo through $^{98}$Mo(n,$\gamma$) reaction is found to be 132.05$\pm$32.89 mb using the activation method and off-line $\gamma$-ray spectrometric technique \cite{badwar}.

The suggested nuclei ($^{95-98}$Mo) in the present work are stable isotopes of Mo. Since the $\beta$ decay lifetime of $^{95-99}$Mo is temperature-sensitive, one must also take into account the fact that the production and destruction of $^{95-99}$Mo depend on the $^{22}$Ne($\alpha$,n)$^{25}$Mg reaction rate (act as neutron sourse), which is even more temperature-sensitive. The race between $^{95-99}$Mo(n,$\gamma$)$^{96-100}$Mo and $^{95-99}$Mo $\rightarrow$  $^{95-99}$Tc + $e$ + $\nu_e$ determines the abundance of $^{95-99}$Mo in the $s$-process environment. Therefore, higher temperatures do not always result in lower Mo abundances. The increase in neutron production rates from the $^{22}$Ne($\alpha$,n)$^{25}$Mg reaction at high temperatures more than compensates for the thermally enhanced $\beta^{-}$ decay rate.

The present study is structured as follows: Section 2 provides a quick overview of the basic formalism.  Section 3 summarizes the results of our calculations and how they compare to previous findings and measurements. Our findings are summarized in Section 4.


\section{Theoretical Framework}
\subsection{The Talys v1.96 code}
The Talys v1.96 code \cite{TALYS1.96} is based on the Hauser-Feshbach theory~\cite{hf1, hf2}. The main inputs in the Hauser-Feshbach theory are the NLDs, the optical model potentials (OMPs), and the gamma strength functions (GSFs). The effect of altering the OMPs can be disregarded in favor of the other two components when low-energy neutrons are used as incident particles (NLDs and GSFs) \cite{hfsr}. The optical model, employed in this study, is the local OMP by Koning and Delaroche \cite{kd}. The results were evaluated for all combinations of the phenomenological NLDs (constant temperature model (CTM), back-shifted Fermi gas model (BSFM), generalized superfluid model (GSM) \cite{ctm, bfm, gsm1, gsm2}) and GSFs (Kopecky-Uhl Lorentzian, Brink-Axel Lorentzian, and Gogny D1M QRPA \cite{brink1, brink2, kop, gogny}). To describe NLD at a quantitative level, it is desired that the
model is able to reproduce all experimentally known discrete
levels that are available in a published database. Without this
step, one may doubt the model’s predictive ability for cases
where no data exist. {The nuclear level density (NLD) is a crucial input in statistical computations since it is employed when information about the discrete level is unavailable.}   
The predictions of the Talys v1.96 code have been compared with the  existing experimental data. The MACS have been computed with neutron energies up to 100 keV. The abundances for the $s$- and the $r$- processes may be reproduced through the sensitivity simulations of these processes. However, one of the major sources of uncertainty lies in the correct estimation of nuclear properties such as the neutron capture cross-section and rates for the nuclei involved in these processes.

The Talys v1.96 code used for the simulation of nuclear reactions includes several state-of-the-art nuclear models to cover almost all key reaction mechanisms encountered for light particle-induced nuclear reactions. It provides an extensive range of reaction channels. The possible incident particles can be simulated in the $E_i$=(0.001–200)~MeV energy range, and  the target nuclides {can be} from $A$=12 {onwards}. The output of the nuclear reaction includes fractional and total cross-sections, angular distributions, energy spectra, double-differential spectra, MACS, and capture rates.  Radiative capture is important in the context of nuclear astrophysics in which a projectile {fuses} with the target nucleus and {emits $\gamma$- ray} \cite{Kabir1,Kabir2,Kabir3,Kabir4,Kabir5}. The nuclear cross-section is an important factor in the calculation of radiative capture rates. The Maxwellian averaged cross-section is used when the energies of the projectiles follow a Maxwellian distribution, like in the stellar environment. MACS is an average of the cross-section over a range of energies, weighted by the Maxwell-Boltzmann distribution.
\begin{align}
{\langle \sigma \rangle (kT) = \frac{2}{\sqrt{\pi}(kT)^2}\int_0^\infty E \sigma (E) exp(\frac{-E}{kT}) dE,}
\end{align}
where $k$, is the Boltzmann constant, $T$ is the temperature, $\sigma(E)$ is the capture cross-section and $E$ is the projectile energy.
{In statistical models for predicting nuclear reactions, level densities are needed at excitation energies where discrete level information is not available or {is} incomplete. Together with the optical model potential, a correct level density is perhaps the most crucial ingredient for a reliable theoretical analysis of cross-sections, angular distributions, and other nuclear {quantities}. Six different level density models, among them three phenomenological and the {rest microscopic}, are {included in the Talys code}. For each of the phenomenological models, the constant temperature model (CTM), the back-shifted Fermi (BSFM) gas model, and the generalized superfluid model (GSM), a version without explicit collective enhancement is considered. In the CTM, the excitation energy range is divided into low energy regions i.e., {from 0 keV up to the matching energy ($E_{M}$) and high energy above the $E_{M}$ where the fermi-gas model (FGM) applies}. Accordingly, the constant temperature part of the total level density reads as} 
\begin{align}
	{\rho_{T}^{tot}(E_x) = \frac{1}{T}exp(\frac{E_x-E_o}{T})},
\end{align}\label{aa}
{where $T$ and $E_o$ serve as adjustable parameters in the constant temperature expression.  The BSFM is used for the whole energy range by treating the pairing energy as an adjustable parameter}
\begin{align}
	{\rho_{F}^{tot}(E_x) = \frac{1}{\sqrt{2\pi}\sigma}\frac{\sqrt{\pi}}{12}\frac{exp(2\sqrt{aU})}{a^{1/4}U^{5/4}}},
\end{align}
{where $\sigma$ is the spin cut-off parameter, which represents the width of the angular momentum distribution, $U$ is the  effective excitation energy and $a$ is the level density parameter defined below} 
\begin{equation}\label{aa}
\begin{aligned}
	{a = \tilde{a}(1+\delta W \frac{1-exp(-\gamma U)}{U})}, 
\end{aligned}
\end{equation}
where $\tilde{a} (=\alpha A+\beta A^{2/3})$ is the asymptotic level density without any shell effects. $A$ is the mass number, $\alpha$, $\beta$ and $\gamma$ are global parameters that need to be determined to give the best average level density description over a whole range of nuclides. They are determined by fitting to experimental level density data. $\delta W$ gives the shell correction energy, and the damping parameter $\gamma$ determines how rapidly \textit{$a$} approaches to $\tilde{a}$. One should note that for the best fitting, $a$ can be readjusted to achieve the desired value of cross-section and nuclear reaction rates. For further investigations, one can {refer to}~\cite{koni}. The GSM is similar to CTM in the way that it also divides the energy range into low and high energies. The high energy range is described by BSFM as mentioned before. It is characterized by a phase transition from a superfluid behavior at low energies, where the pairing correlations strongly influence the level density.

\begin{align}
	{	\rho^{tot}(E_x) = \frac{1}{\sqrt{2\pi}\sigma}\frac{e^S}{\sqrt{D}}},
\end{align}
{where $S$ is the entropy and $D$ is the determinant related to the saddle-point approximation.}
{The gamma strength function (GSF) plays a crucial role in estimating cross-sections and reaction rates, particularly in processes involving the emission of gamma rays.  Different GSFs are included in Talys v1.96, among them the Brink-Axel model is used for all transitions except for $E1$ \cite{TALYS1.96}. The GSF function $f_{XL}$ gives the distribution of the average reduced partial transition width as a function of the photon energy $E_{\gamma}$. }
\begin{align}
	{f_{XL}(E_{\gamma}) = K_{XL}\frac{\sigma_{XL}E_{\gamma}\Gamma^2_{XL}}{(E_{\gamma}^2-E_{XL}^2)^2+(E_{\gamma}\Gamma^2_{XL})^2}},
\end{align}
where $E_{XL}$ is the evergy, $\Gamma_{XL}$ is the width, and $\sigma_{XL}$ is the giant resonance strength. For $E1$ transitions, Talys v1.96 utilizes the Kopecky-Uhl model by default.
\begin{equation}
	\begin{aligned}
	f_{XL}(E_{\gamma}, T)= K_{XL}\big[\frac{E_{\gamma}\tilde{\Gamma}_{E1}(E_{\gamma})}{(E_{\gamma}^2-E_{E1}^2)^2+E^2_{\gamma}\tilde{\Gamma}_{E1}(E_{\gamma})^2}+
	\\
	\frac{0.7\Gamma_{E1}4\pi^2T^2}{E_{E1}^3}\big]\sigma_{E1}\Gamma_{E1}
	\end{aligned}
\end{equation}
and
\begin{align}
	\tilde{\Gamma}_{E1}(E_{\gamma}) = \Gamma_{E1}\frac{E_{\gamma}^2+4\pi^2}{E_{E1}^2}\big[\frac{E_n+S_n-\Delta-E_{\gamma}}{a(S_n)}\big],
\end{align}
{where $\tilde{\Gamma}(E_{\gamma})$ represents the energy-dependent damping width, $E_n$ is the incident energy of neutrons, $S_n$ is the neutron seperation energy, and $\Delta$ is the correction for pairing.  $a$ represents the level density parameter at $S_n$ as mentioned earlier.}

\subsection{\textbf{The pn-QRPA model}}

The stellar $\beta$ -decay rates are investigated within the framework of  pn-QRPA. The
Hamiltonian of the model is presented as
\begin{equation}
	H^{QRPA} = H^{sp} + V^{pair} + V ^{ph}_{GT} + V^{pp}_{GT}.
	\label{Eqt. Hamiltonian}
\end{equation}
The deformed Nilsson potential ($H^{sp}$) basis was utilized to compute wave functions and single particle energies. $V^{pair}$ pairing forces were treated employing the BCS formalism. {Q-values and residual interactions have a considerable influence on the computed electron emission ($\beta^{-}$) rates and associated half-lives
	\cite{Engel99}.} The $V ^{ph}_{GT}$ ($\chi$ ($\textit{ph}$)), $V^{pp}_{GT}$ ($\kappa$ ($\textit{pp}$)) known as the residual interactions, {were taken into account for the calculation of GT strength.} For a thorough definition of $\chi$ and $\kappa$, as well as the optimal choice of these parameters, see into Refs.~\cite{Engel99,Sta90,Hir93,Hom96}. 

{The  $V^{ph}_{GT}$ force was found employing
	\begin{equation}\label{ph}
		V^{ph}_{GT}= +2\chi\sum^{1}_{\mu= -1}(-1)^{\mu}Y_{\mu}Y^{\dagger}_{-\mu},
	\end{equation}
	with
	\begin{equation}\label{y}
		Y_{\mu}= \sum_{j_{p}m_{p}j_{n}m_{n}}<j_{p}m_{p}\mid
		t_- ~\sigma_{\mu}\mid
		j_{n}m_{n}>c^{\dagger}_{j_{p}m_{p}}c_{j_{n}m_{n}}.
	\end{equation}
	Whereas the  $V^{pp}_{GT}$ interaction was determined employing
	\begin{equation}\label{pp}
		V^{pp}_{GT}= -2\kappa\sum^{1}_{\mu=
			-1}(-1)^{\mu}P^{\dagger}_{\mu}P_{-\mu},
	\end{equation}
	with
	\begin{eqnarray}\label{p}
		\begin{aligned}
			P^{\dagger}_{\mu}= \sum_{j_{p}m_{p}j_{n}m_{n}}<j_{n}m_{n}\mid
			(t_- \sigma_{\mu})^{\dagger}\mid
			j_{p}m_{p}>\\
			\times (-1)^{l_{n}+j_{n}-m_{n}}c^{\dagger}_{j_{p}m_{p}}c^{\dagger}_{j_{n}-m_{n}},
		\end{aligned}
	\end{eqnarray}
where the operators are explained below and the other symbols have their normal meanings. The $\chi$ and $\kappa$ were taken from Ref.~\cite{Hom96}. Reduced GT transition probabilities were achieved by expressing the QRPA ground state into one-phonon states in the daughter nucleus. Additional input variables for the calculation of weak transitions include the pairing gap ($\Delta_{p}$, $\Delta_{n}$), nuclear deformation ($\beta_{2}$), threshold-values of energy and Nilsson potential variables
(NPV).  We adjusted our computation with nuclear deformation parameters from the most current analysis \cite{Mol2012}. The NPV were taken from \cite{Rag84}, and the oscillator constant (which is similar for proton and neutron) was identified using the equation $\hbar\omega=41A^{-{1}/{3}}$ (in MeV). We employed the Nilsson potential to calculate the wave function. The Nilsson framework is often used to characterize the framework of low-lying states. We used the  $\beta_{2}$ as a Nilsson potential input variable. Primarily, wave functions and single particle energies were calculated on the deformed Nilsson basis. The transformation from the spherical nucleon basis (\textit{c}$_{jm}^{+}$, \textit{c}$_{jm}$) to the axial symmetric deformed basis
	(\textit{d}$_{m\alpha}^{+}$, \textit{d}$_{m\alpha}$) were performed
	\begin{equation}\label{ND}
		d_{m\alpha}^{+}=\sum_{j}D_{j}^{m\alpha}c_{jm}^{+}.
	\end{equation}
	The \textit{D}$_{j}^{m\alpha}$ is a group of Nilsson eigenfunctions with $\alpha$ as an additional quantum number to characterize the Nilsson eigen-states. The BCS formalism was used in the Nilsson basis for neutron/proton system separately. The diagonalization of the Nilsson Hamiltonian yielded the transformation matrices (a detailed explanation may be found in \cite{Hir93, Mut89}).
	The globally systematic pairing gap values, $\Delta_{n}$=$\Delta_{p}$=12$/\sqrt{A}$ (MeV), were employed in our computation. The Q-values were obtained from the most current assessment of atomic mass data by \cite{Audi2017}.
	Further details of solution of  Eq.~(\ref{Eqt. Hamiltonian}) may be studied from \cite{Mut92}. One may view the computation of terrestrial $\beta^{-}$ half-lives in Ref. \cite{Sta90}. Further information on the formalism utilized to estimate GT transitions in stellar scenarios using the pn-QRPA technique can be found in Refs. \cite{Nabi99c,Nabi04}.
	
	The electron emission rates/positron capture ($pc$) rates ($\lambda^{(\beta^{-}/pc)}_{ij}$) from the parent nucleus ($\mathit{i}$th-state) to the daughter nucleus ($\mathit{j}$th-state) is given by
	\begin{equation}\label{EE}
		\lambda^{^{(\beta^{-}/pc)}}_{ij}=ln2\frac{f_{ij}(T,\rho,E_f)}{(ft)_{ij}},
	\end{equation}
	where $(ft)_{ij}$ is related to the reduced transition probability (B$_{ij}$) by the relation
	\begin{equation}\label{ft}
		(ft)_{ij}=D/B_{ij}.
	\end{equation}
	The value of $D$ is taken as $6143$~s \cite{Hardy}, and $B_{ij}$ is defined below
	\begin{equation}\label{BGT}
		B_{ij}=B(F)_{ij}+(g_A/g_{V})^2B(GT)_{ij},
	\end{equation}
	where B(F) and B(GT) are reduced transition probabilities of the Fermi and Gamow Teller transitions, respectively. $f_{ij}$ are the phase space integrals. Further details can be found in Ref.~\cite{Nabi04}.  Because of the high temperatures in the stellar core, $\beta^{-}$ and $pc$ rates get only a minor contribution from parent excited energy levels. We utilize the Boltzmann distribution function to compute the occupancy probability of the parent $ith$-state
	\begin{equation}\label{pi}
		P_{i} = \frac {exp(-E_{i}/kT)}{\sum_{i=1}exp(-E_{i}/kT)}.
	\end{equation}
	Furthermore, the total stellar $\beta^{-}$/$pc$ rates were computed using
	\begin{equation}\label{lb}
		\lambda^{^{(\beta^{-}/pc)}} = \sum_{ij}P_{i} \lambda_{ij}^{^{(\beta^{-}/pc)}}.
	\end{equation}
	The summation stands for computation of all parent and daughter
	energy levels until satisfactory convergence is achieved. We noticed that in our present work of weak $\beta^{-}$ and $pc$ calculations, a large model space (up to 7~$\hbar\omega$ major oscillatory shells) makes it easier to achieve the desired convergence. The ability to calculate the weak rates of any heavy nuclear species is one of the main advantages of the pn-QRPA technique with the chosen Hamiltonian (Eq.~(\ref{Eqt. Hamiltonian})).
	\subsection{\textbf{The stellar neutron capture rates ($\lambda_{(n,\gamma)}$)} }
	The total neutron capture rate ($\lambda_{(n,\gamma)}$ in the units of $s^{-1}$) is defined as \cite{43}
	\begin{equation}\label{ng}
		\lambda_{(n,\gamma)} = v_{i}\times \sigma_{i} \times n_n,
	\end{equation}
	where $v_{i}$ is the averaged velocity of neutron, $\sigma_{i}$ is the MACS, and $n_n$ is the average neutron density for T$_{9}$ $<$ 1
	\begin{align}\label{nd}
		n_n = \frac{4.3\times10^{36}\rho X_4e^{[-(0.197/T_9)^{4.82}]}(\frac{1}{T_9})^{2/3} e^{\frac{-47}{(T_9)^{1/3}}}}{\sigma_{22}[1+{\sigma_{i}N_i}/\Sigma\sigma_{22}N_{22}](\frac{T_9}{0.348})^{1/2}},
	\end{align}
where $X_4$ is the helium mass fraction, $\rho$ is the nuclear matter density and T$_9$ is the temperature at the base of the convective shell. $\sigma_{22}$ is the $^{22}$Ne(n, $\gamma$)$^{23}$Ne capture cross-section. $N_i$ is the abundance of species $i$ and  $\sigma_{i}$ is their MACS. The composition of heavy nuclei at the base of the convective shell is dominated by $^{22}$Ne produced by $\alpha$- captures on $^{14}$N from the CNO cycle.

\section{Results and Discussion} 
For some applications, such as astrophysical investigations involving nuclei along neutron or proton drip lines, it is crucial to extrapolate the data much beyond the experimentally known region. Large scale applications thus need to use data from reliable theoretical models. Inaccurate NLD description or prediction is the main source of uncertainty in statistical model predictions. Since experiments are not attainable at all energies, it is evident that theoretical extrapolations are critical. Here, we have used the nuclear level density model and a variety $\gamma$-strength functions to compute the MACS for $^{95-98}$Mo isotopes. The present-listed Mo isotopes are almost $\beta$ stable in terrestrial environments.  We have investigated the MACS $^{95-98}$Mo(n,$\gamma$)$^{96-99}$Mo for a range of temperatures appropriate for the presumed site of the $s$-process of nucleosynthesis.
\begin{figure}
	\centering
	\includegraphics[width=0.82\textwidth]{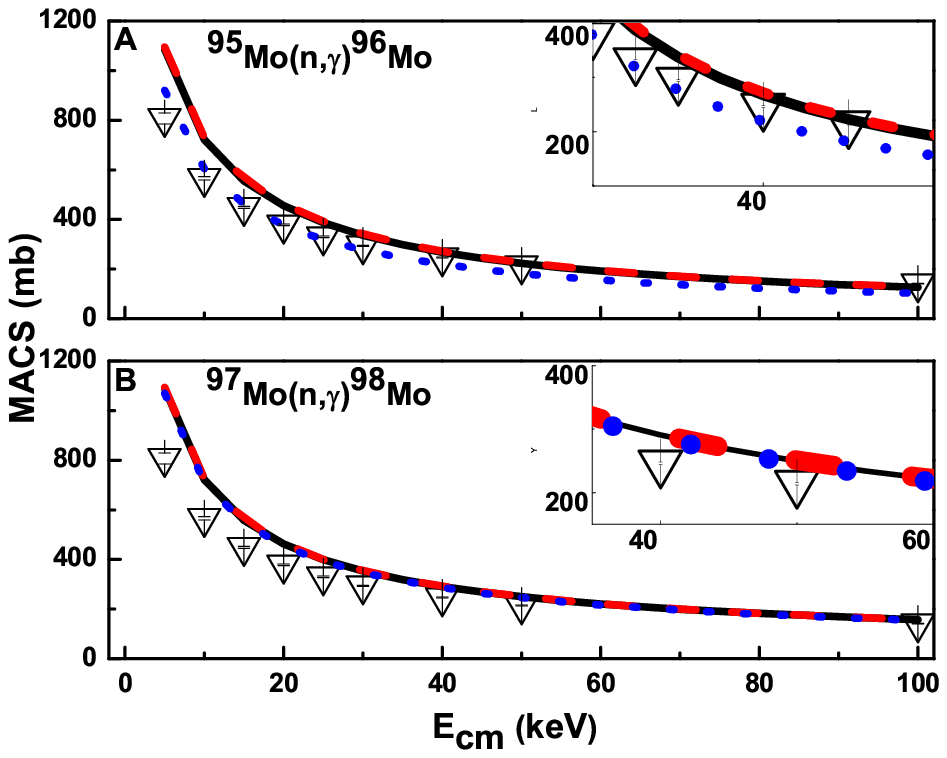}
	\vspace*{-43mm}
	\caption{{{{The total MACS for $^{95,97}$Mo(n,$\gamma$)$^{96,98}$Mo along with the measured data ($\bigtriangledown$) \cite{Winters1}. (A) The computed MACS for the $^{95}$Mo(n,$\gamma$)$^{96}$Mo by the BSFM with the GSFs as Brink-Axel (solid line), {Gogny (dashed line)} and {Kopecky-Uhl} (dotted line). (B) The computed MACS for the $^{97}$Mo(n,$\gamma$)$^{98}$Mo by the BSFM with the GSFs as Brink-Axel (solid line), {Gogny (dashed line)} and {Kopecky-Uhl} (dotted line).}}}}
	\label{fig:1}
\end{figure}
The consideered isotopes under investigations have almost magic neutron configurations, which results in their small (n,$\gamma$) cross-sections and fairly large $s$-abundances. The MACS  $^{95-98}$Mo(n,$\gamma$)$^{96-99}$Mo are computed via Talys v1.96 in the Hauser-Feshbach framework by choosing the NLDs, and a radiative strength function. Out of several possibilities for level density calculations, we employed the phenomenological NLDs as BSFM for excitation energies up to 100~keV. The obtained NLDs computed by the BSFM were employed for the calculation of MACS for   $^{95-97}$Mo(n,$\gamma$)$^{96-98}$Mo. One should note that we emplyed NLD as BSFM and varied the GSFs. For example, we have fixed the NLD as a BSFM and varied the GSFs as Brink-Axel, Gogny, and Kopecky-Uhl.
\begin{figure}
	\centering
	\includegraphics[width=0.82\textwidth]{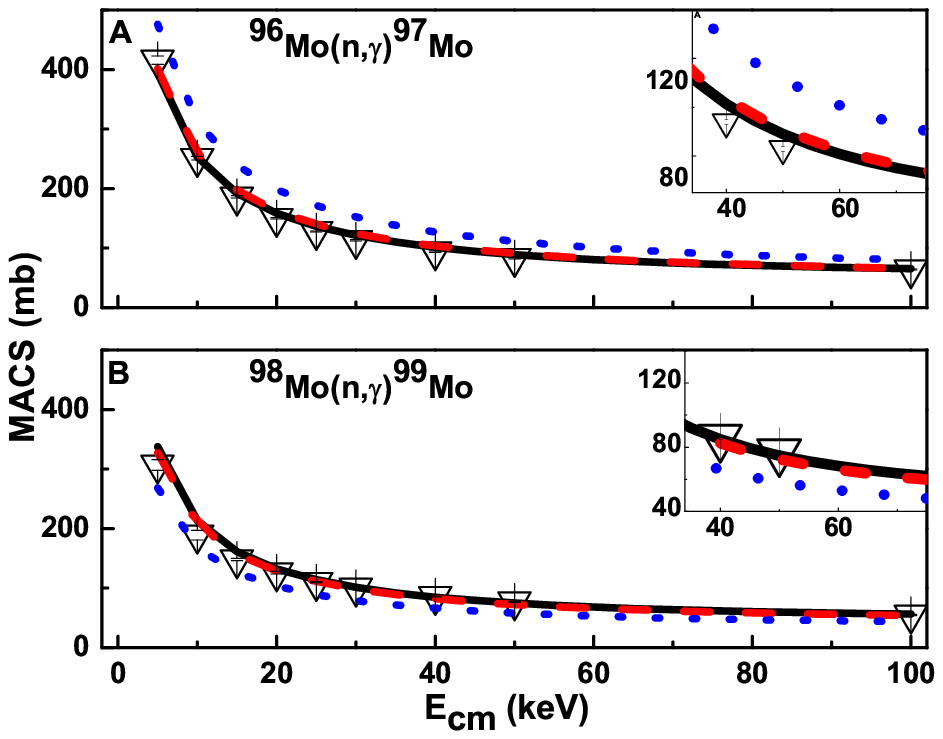}
	\vspace*{-43mm}
	\caption{{{The total MACS for $^{96,98}$Mo(n,$\gamma$)$^{97,99}$Mo along with the measured data ($\bigtriangledown$) \cite{Winters1}. (A) The computed MACS for the $^{96}$Mo(n,$\gamma$)$^{98}$Mo by the BSFM with the GSFs as Brink-Axel (solid line), {Gogny (dashed line)} and {Kopecky-Uhl} (dotted line). (B) The computed MACS for the $^{96}$Mo(n,$\gamma$)$^{99}$Mo by the BSFM with the GSFs as Brink-Axel (solid line), {Gogny (dashed line)} and {Kopecky-Uhl} (dotted line).}}}
	\label{fig:2}
\end{figure}
Together with the findings of Ref.~\cite{Winters1}, the Maxwellian average cross-section results for $^{95, 97}$Mo(n,$\gamma$) within (0.01$<$$kT$$<$100 keV) are shown in Figs.~(\ref{fig:1})A,B. The computed results based on the present model agree well with the data of Ref.~\cite{Winters1}, both above and below the $s$-process energy $kT$=30~keV. Note that in order to get the best-fitting with the measured data, we adjusted the parameter $a$ as mentioned in Eq.~(\ref{aa}). In the present case, the  adjustment parameter $a$ were obtained 0.5 through adjustment to all available experimental of the MACS.  Similarly, $^{96,98}$Mo(n,$\gamma$)$^{97,99}$Mo radiative capture processes were analyzed using the NLD as a BSFM and the same sets of GSFs (as previously mentioned). Fixing the NLD as BSFM and changing the GSFs as Gogny, Kopecky-Uhl, and Brink-Axel. The energy dependence Maxwellian averaged cross-sections for $^{96}$Mo and $^{98}$Mo isotopes are depicted in Figs.~(\ref{fig:2})A,B along with the results of Ref.~\cite{Winters1}. The computed MACS at $kT$=30 keV in the analysis of $^{96}$Mo(n,$\gamma$)$^{97}$Mo were 121.35 mb (GSF as Brink-Axel), 124.31 mb (GSF as Gogny), and 152.29 mb (GSF as Kopecky-Uhl). It is observed that the current model-based results agree well with the 113~mb MACS at $kT$=30 keV found in Ref.~\cite{Winters1}. Similarly, the computed MACS at $kT$=30 keV for the analysis of $^{98}$Mo(n,$\gamma$)$^{99}$Mo were 101.28 mb (GSF as Brink-Axel), 98.42 mb (GSF as Gogny), and 78.80 mb (GSF as Kopecky-Uhl). Remarkably, at $kT$=30 keV, the current model-based results agree well with the MACS reported in Ref.~\cite{Winters1}. 
 
\begin{figure}
	\centering
	\includegraphics[width=0.8\textwidth]{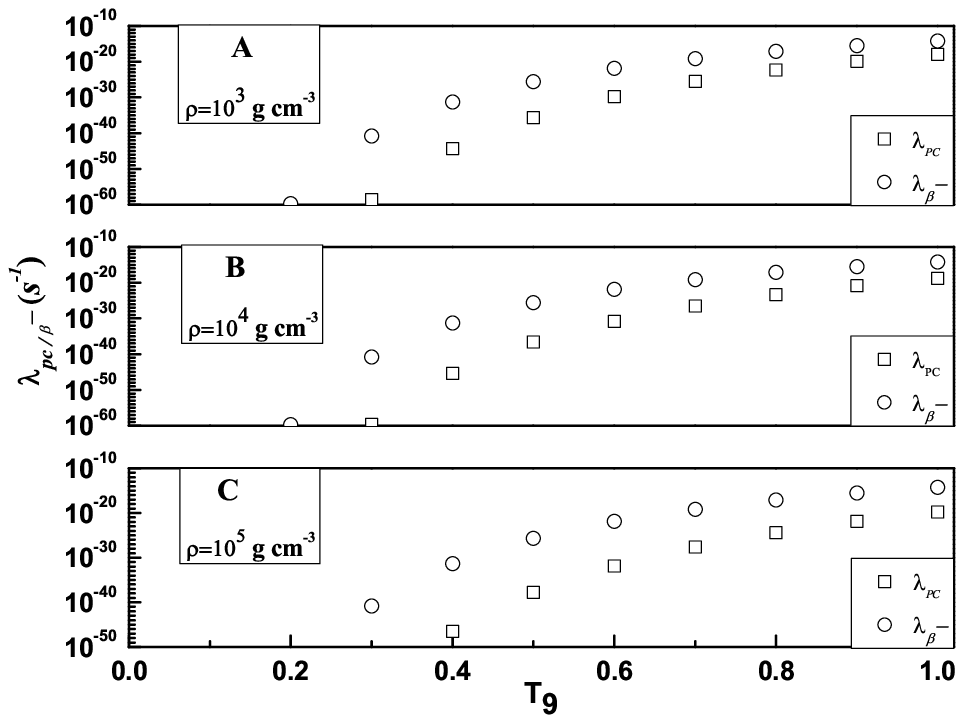}
	\vspace*{-47mm}
 	 	\caption{{{{The $\lambda_{pc}$ and  $\lambda_{\beta^{-}}$ rates for the $\beta$ decay of $^{95}$Mo.}}}}
	\label{fig:3}
\end{figure}
Stellar $\beta$ rates of $^{95-98}$Mo nuclei have been analyzed using the pn-QRPA model in the second part of our study. The pn-QRPA Hamiltonian  defined by Eq.~(\ref{Eqt. Hamiltonian}) has Nilsson deformed potential and residual interaction terms. In the present study, we used the nuclear deformation ($\beta_{2}$) from the Finite Range Droplet Model (FRDM) model as an input parameter in our pn-QRPA model for the selected Mo isotopes. The nuclear $\beta_{2}$ value was used in our computation as a free parameter. 
 \begin{figure}[h!]
 	\centering
 	\includegraphics[width=0.8\textwidth]{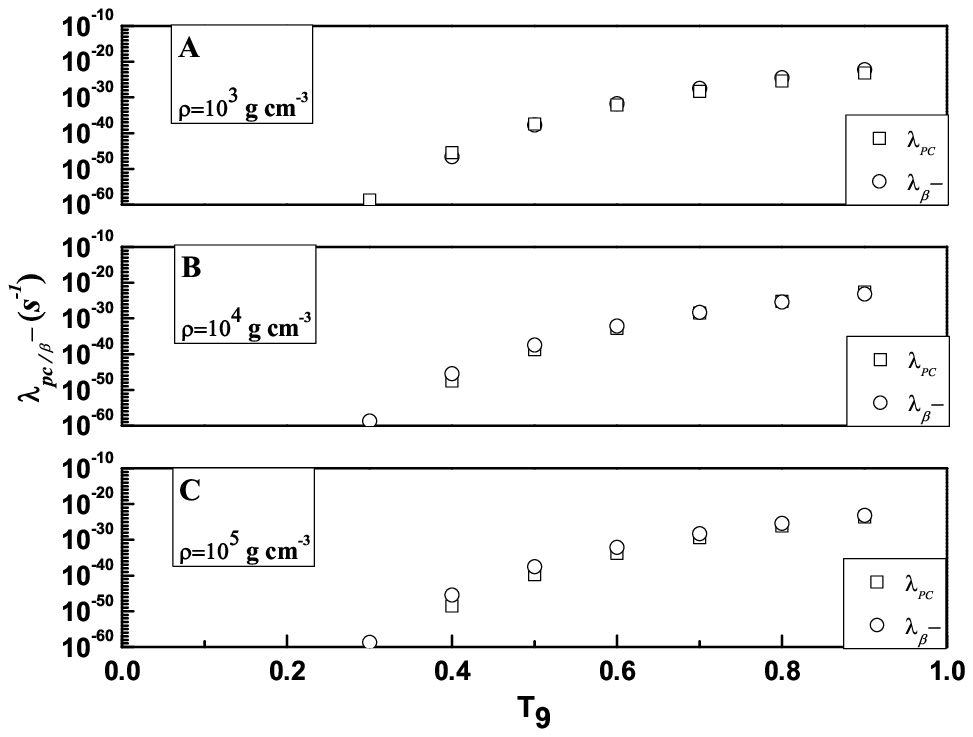}
 	\vspace*{-48mm}
  	\caption{{{{The $\lambda_{pc}$ and  $\lambda_{\beta^{-}}$ rates  for the $\beta$ decay of $^{96}$Mo.}}}}
 	\label{fig:4}
 \end{figure}
The $\beta_{2}$ values from the most recent globally updated FRDM \cite{Mol2012} computation served as the foundation for our calculations. The $\beta_{2}$ values are 0.000 ($^{95}$Mo), 0.150 ($^{96}$Mo), 0.172 ($^{97}$Mo) and 0.206 ($^{98}$Mo). The residual interactions $\chi$ and $\kappa$ values in the present cases are optimized based on Ref.~\cite{Hom96} and satisfied the Ikeda sum rule. Based on the optimized values of the residual interactions, we have computed the stellar $\beta^{-}$ decay and $pc$ rates.  The results of our present investigations are depicted in Figs.~(\ref{fig:3}-\ref{fig:6})A-C at densities $\rho$=10$^{3}$-10$^{5}$ g cm$^{-3}$. 
\begin{figure}[h!]
	\centering
	\includegraphics[width=0.8\textwidth]{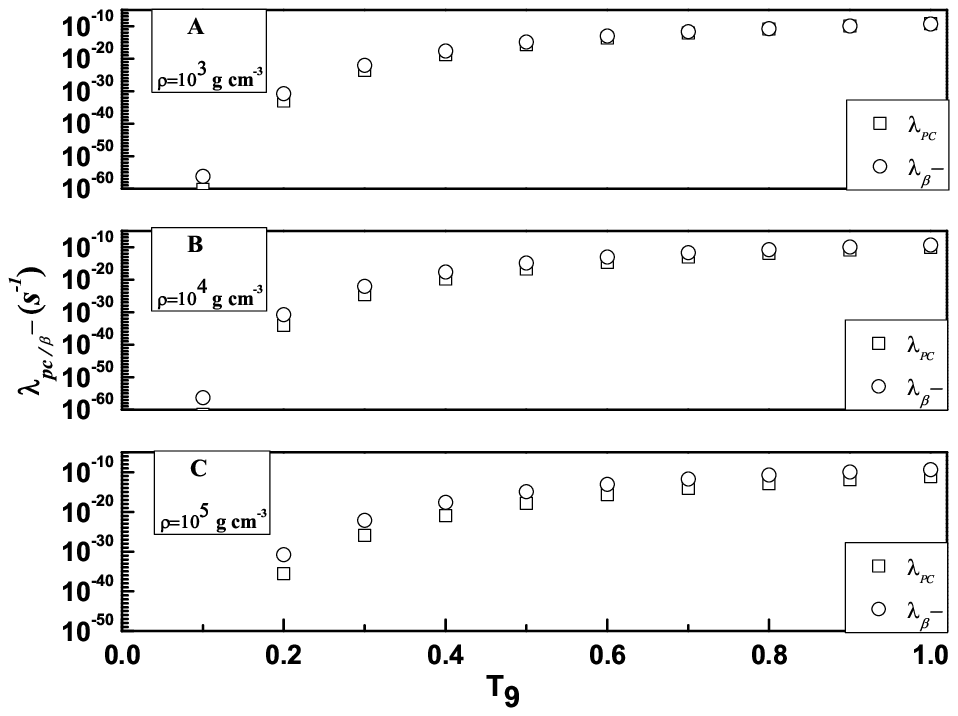}
	\vspace*{-48mm}
	 	\caption{{{{The $\lambda_{pc}$ and  $\lambda_{\beta^{-}}$ rates for the $\beta$ decay of $^{97}$Mo.}}}}
	\label{fig:5}
\end{figure}
Even though these Mo isotopes are stable in terrestrial environments, they have higher $\beta^{-}$ decay than $pc$ rates at lower and even higher temperatures. At low density $\rho$=10$^{3}$ g cm$^{-3}$, and lower temperatures, the $\lambda_{\beta^{-}}$ are higher than the $\lambda_{pc}$; however, at high temperatures, the $\lambda_{pc}$ increases gradually because more positrons are created at high temperatures. For $^{95}$Mo there is a clear difference between the $\lambda_{pc}$ and $\lambda_{\beta^{-}}$ at higher densities, as mentioned in Fig.~(\ref{fig:3})A-C. 
\begin{figure}[h!]
	\centering
	\includegraphics[width=0.8\textwidth]{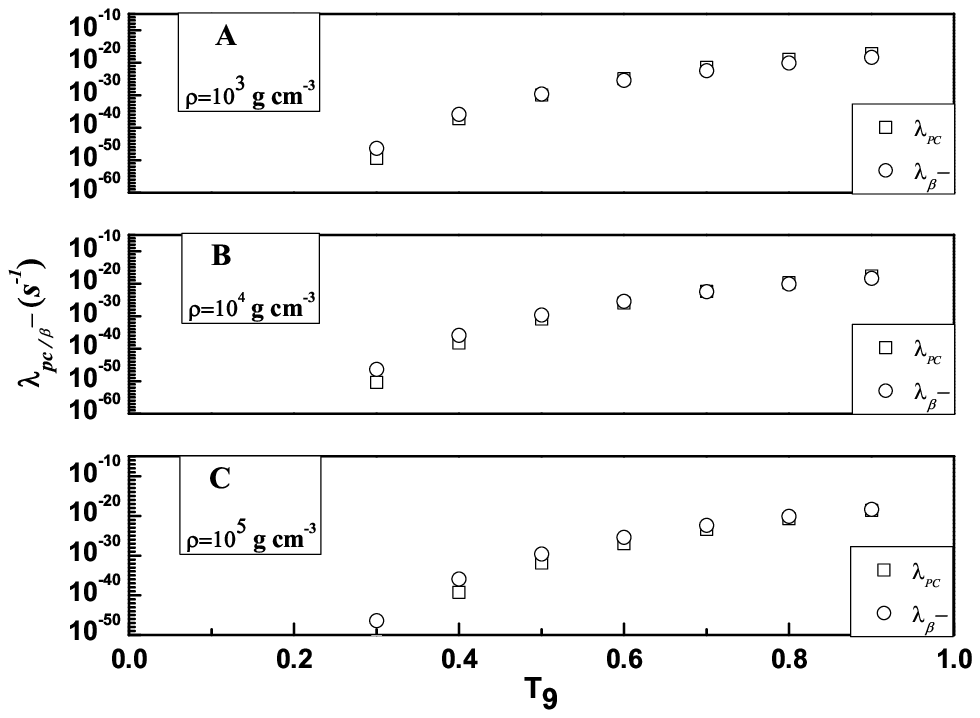}
	\vspace*{-48mm}
 	\caption{{{{The $\lambda_{pc}$, and  $\lambda_{\beta}$ rates, at all densities, for the $\beta$ decay of $^{98}$Mo.}}}}
	\label{fig:6}
\end{figure}
Similarly, Fig.~(\ref{fig:4}-\ref{fig:6})A-C display the $\lambda_{pc}$ and $\lambda_{\beta^{-}}$ rates for $^{96}$Mo\,--$^{98}$Mo nuclei at $\rho$=(10$^{3}$-10$^{5}$) g cm$^{-3}$. We noted that the stellar $\beta^{-}$ decay rates increase with temperature, but in the heavier nuclei of Mo, the $\lambda_{pc}$ and $\lambda_{\beta^{-}}$ are almost the same at higher temperatures.

\begin{figure*}[h!]
	\begin{multicols}{2}
		\includegraphics[width=1.5\linewidth]{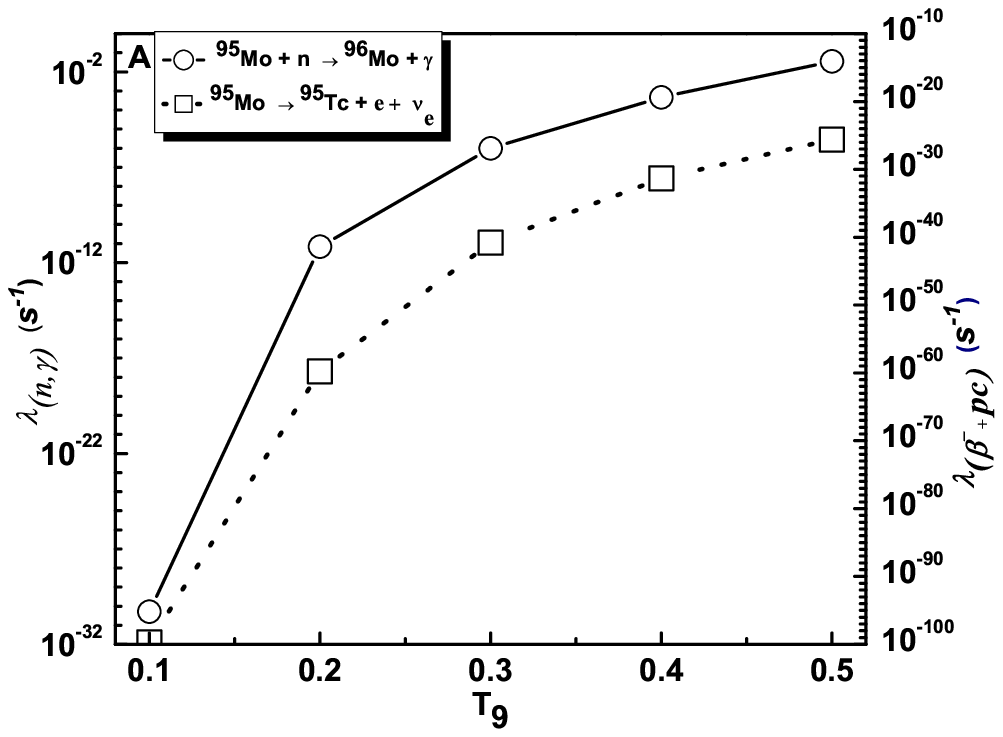}\par 
		\includegraphics[width=1.5\linewidth]{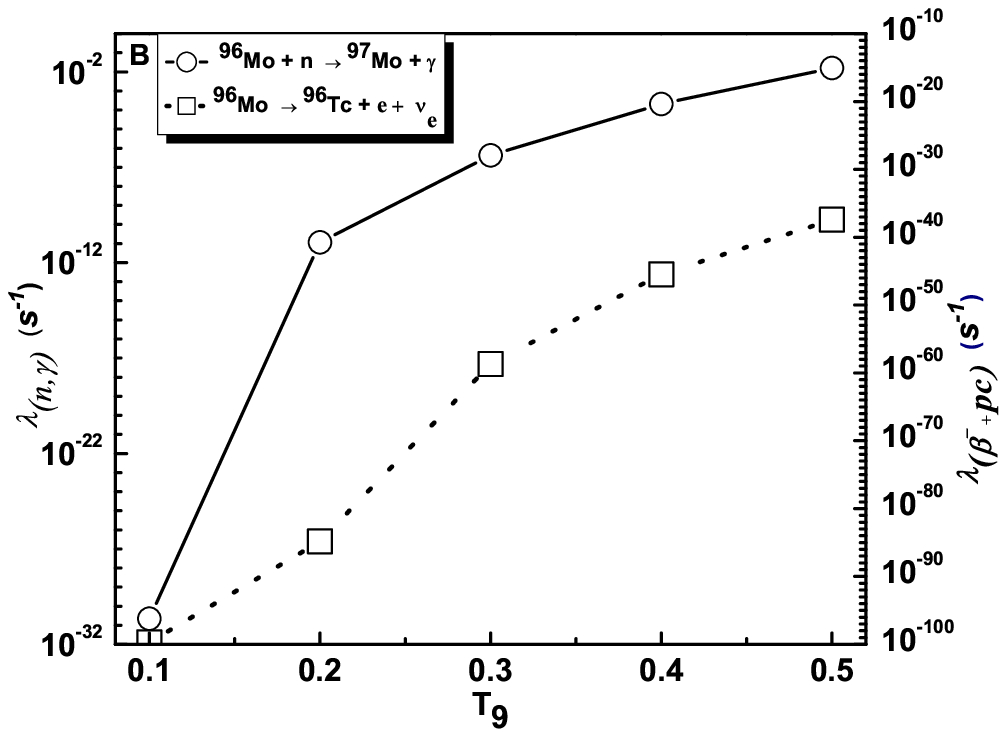}\par 
	\end{multicols}
	\vspace*{-40mm}
	\begin{multicols}{2}
		\includegraphics[width=1.5\linewidth]{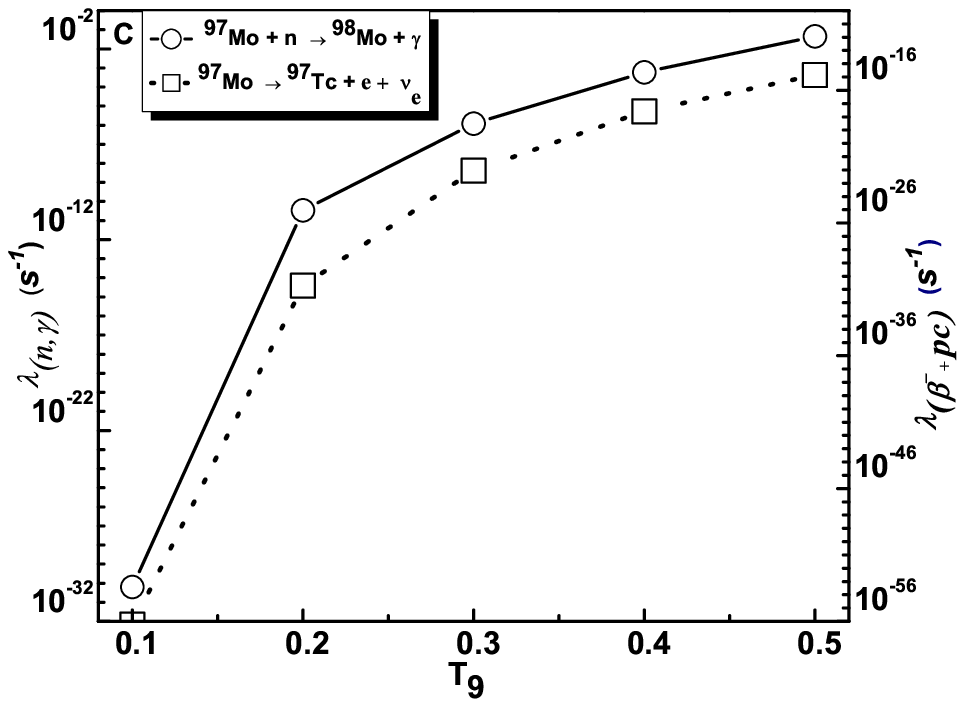}\par
		\includegraphics[width=1.5\linewidth]{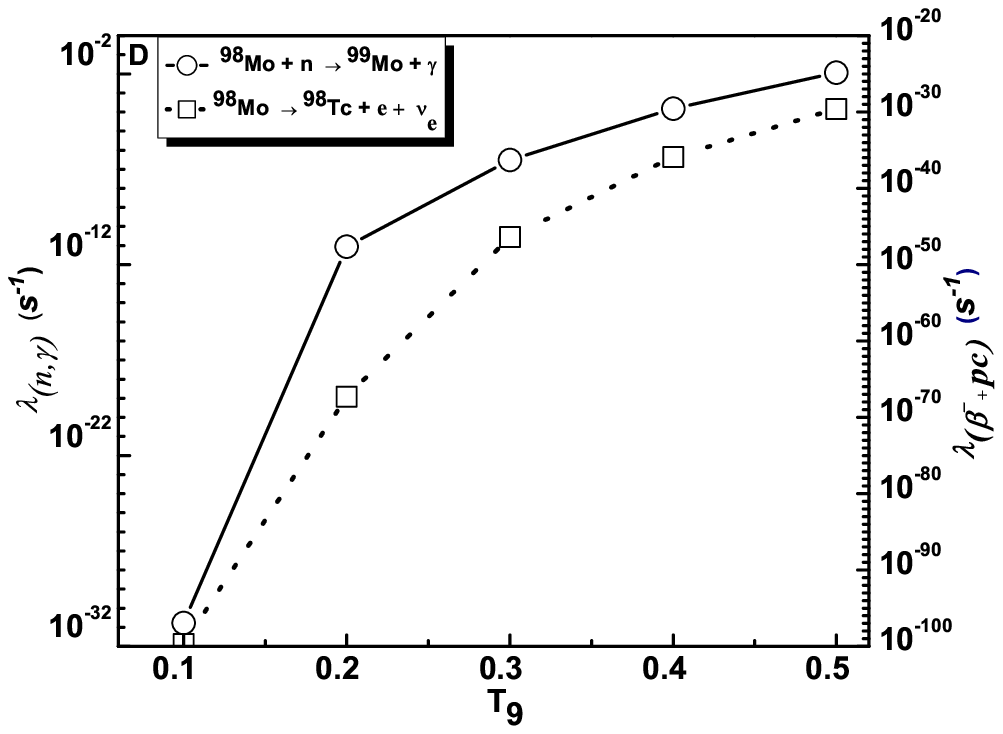}\par
	\end{multicols}
	\vspace*{-44mm}
	\centering\caption{Neutron capture rates ($\lambda_{(n,\gamma)}$) and stellar $\beta$ decay rates ($\lambda_{(\beta^{-}+pc)}$) of $^{95-98}$Mo nuclei at $\rho$=10$^{4}$  g cm$^{-3}$. (A). The computed  $\lambda_{(n,\gamma)}$ and $\lambda_{(\beta^{-}+pc)}$ for  $^{95}$Mo. (B). The computed  $\lambda_{(n,\gamma)}$ and $\lambda_{(\beta^{-}+pc)}$ for  $^{96}$Mo. (C). The computed  $\lambda_{(n,\gamma)}$ and $\lambda_{(\beta^{-}+pc)}$ for  $^{97}$Mo. (D). The computed  $\lambda_{(n,\gamma)}$ and $\lambda_{(\beta^{-}+pc)}$ for  $^{98}$Mo. }
	\label{fig:07}
\end{figure*}

As part of the third phase of this investigation, we have calculated the sum of stellar $\beta^{-}$ decay and $pc$ rates ($\lambda_{(\beta^{-}+pc)}$)  and the temperature-dependent neutron capture rates ($\lambda_{(n,\gamma)}$) for the selected Mo isotopes. For the calculations of $\lambda_{(n,\gamma)}$, we employed Eq.~(\ref{ng}). In the initial set of computations, we identified the best-fitting model combination for each isotope, which in our investigation is Brink-Axel as a GSF and BSFM as an NLD. Based on the relevant MACS, we have computed the $\lambda_{(n,\gamma)}$. 
The present computed MACS for $^{95}$Mo (276 mb at E$_n$=30 keV) and for $^{97}$Mo (347 mb at E$_n$=30 keV). With these cross-sections, the total neutron capture rate for $^{95}$Mo and $^{97}$Mo are $\lambda_{(n,\gamma)}$=6.60$\times$10$^{-17}$$n_n$ and $\lambda_{(n,\gamma)}$=8.305$\times$10$^{-17}$$n_n$, respectively. Similarly, with neutron capture cross-sections for $^{96}$Mo (121 mb at E$_n$=30 keV) and  for $^{98}$Mo (98 mb at E$_n$=30 keV), the total capture rates for $^{96}$Mo and $^{98}$Mo are $\lambda_{(n,\gamma)}$=2.90$\times$10$^{-17}$$n_n$ and $\lambda_{(n,\gamma)}$=2.34$\times$10$^{-17}$$n_n$, respectively. At  $\rho$=10$^{4}$ g cm$^{-3}$ and $X_4$=0.2 the present computed $\lambda_{(n,\gamma)}$ and $\lambda_{(\beta^{-}\,+\,pc)}$ are depicted in Fig.~(\ref{fig:07})A-D for the $^{95-98}$Mo + n $\rightarrow$ $^{96-99}$Mo + $\gamma$ and $^{95-98}$Mo $\rightarrow$ $^{95-98}$Tc + $e$ + $\nu_e$. It is obvious that, at lower temperature the thermally enhanced $\beta$ decay rates are much lower in magnitude than $\lambda_{(n,\gamma)}$. For example, at T$_{9}$=0.1, $\lambda_{(n,\gamma)}$=5.1$\times$10$^{-31}$ s$^{-1}$ while at the same temperature, $\lambda_{(\beta^{-}\,+\,pc)}$ is 0.97$\times$10$^{-100}$ s$^{-1}$ for $^{95}$Mo. At higher temperature T$_{9}$=0.5, $\lambda_{(n,\gamma)}$=3.5$\times$10$^{-02}$ s$^{-1}$ and  $\lambda_{(\beta^{-}\,+\,pc)}$ is 2.37$\times$10$^{-26}$ s$^{-1}$. For $^{97}$Mo, at T$_{9}$=0.1,  $\lambda_{(n,\gamma)}$=6.4$\times$10$^{-31}$ s$^{-1}$ while at the same temperature $\lambda_{(\beta^{-}+pc)}$ is 5.5$\times$10$^{-56}$ s$^{-1}$. But at T$_{9}$=0.5  $\lambda_{(n,\gamma)}$=4.4$\times$10$^{-02}$ s$^{-1}$ and  $\lambda_{(\beta^{-}+pc)}$ is 1.37$\times$10$^{-15}$ s$^{-1}$. Therefore, in the examined physical conditions, the neutron capture rates of Mo isotopes exceed the thermally enhanced beta decay rates both at lower and higher temperatures.

\section{Conclusion}
Below we present a summary of our main findings and make some inferences.\\
1. We have examined the MACSs of the ($n$,$\gamma$) process for the sets of $\beta$ stable Mo nuclei with magic or almost magic neutron numbers in the context of the Talys v1.96 {code} across a broad energy range. We deduced lower capture cross-section for $^{96,98}$Mo(n,$\gamma$)$^{97,99}$Mo than $^{95,97}$Mo(n,$\gamma$)$^{96,98}$Mo, usually included in models of the $s$-process in AGB stars. It should be noted that Talys provides a data representation of the phenomenological sets of NLDs and GSFs that are relatively similar. Several NLD and GSF models from Talys have been used to generate the MACS for the $^{95-98}$Mo(n,$\gamma$)$^{96-99}$Mo processes. Following that, the generated values were contrasted with the measurements found in the literature. We found that, out of all the GSF models, the Brink-Axel of Talys is the best fit for the $^{95-98}$Mo(n,$\gamma$)$^{96-99}$Mo processes. For these nuclei, the Hauser-Feshbach theory prediction accurately reproduces the available experimental data with all parameter modifications and model variations. \\
2. We computed the $\beta^{-}$/$pc$ rates based on the pn-QRPA framework for different densities and temperatures.  The pn-QRPA theory effectively calculates weak interaction rates for $^{96-98}$Mo. Our analysis included a broad model space of 7~$\hbar\omega$.   Weak rates for a wide range of stellar temperatures and densities were calculated.  We noted that the weak rates increase directly with temperature. This is due to the fact that at higher temperatures, the occupation probability of parent excited states increase and partial rates from parent excited levels have a finite contribution to the total rate~(Eq.~(\ref{lb})).\\ 
3. We compared $\lambda_{(n,\gamma)}$ and $\lambda_{(\beta^{-}+pc)}$ under an appropriate stellar scenario in the third phase of our investigations. It was noted that the stellar $\beta$ -decay rates are much lower than the neutron capture rates at low temperatures. At higher temperatures, the stellar $\beta$ decay increases but still are much smaller than the $\lambda_{(n,\gamma)}$ rates. For example, at T$_{9}$=0.5, the $\lambda_{(n,\gamma)}$/$\lambda_{(\beta^{-}+pc)}$=1.50$\times$10$^{24}$, 4.27$\times$10$^{35}$, 3.22$\times$10$^{13}$ and 4.48$\times$10$^{27}$ for $^{95-98}$Mo, respectively.

From the present study, we concluded that the $\lambda_{(n,\gamma)}$ are higher than $\lambda_{(\beta^{-}+pc)}$ rates. 
\pagebreak
\section{References}

\end{document}